\documentclass[floatfix,showpacs,aps,nofootinbib]{revtex4}
\usepackage{amssymb}

\usepackage[dvips]{epsfig}
\usepackage[english]{babel}

\usepackage{bbm}
\usepackage[utf8]{inputenc}
\usepackage{array}
\usepackage{amsmath}
\usepackage{amsfonts}
\usepackage{amssymb}
\usepackage{xcolor}
\usepackage{hyperref}
\hypersetup{colorlinks=true,linkbordercolor=blue,linkcolor=blue, citecolor=blue}
\usepackage{subeqnarray}
\usepackage{tensor}
\usepackage{indentfirst} 

\makeatletter
\renewcommand*\l@section{\@dottedtocline{1}{1.5em}{2.3em}}
\makeatother

\begin{document}

\title{Non-standard Dirac adjoint spinor: The emergence of a new dual}
\author{R. J. Bueno Rogerio$^{1}$}\email{rodolforogerio@feg.unesp.br}
\author{C. H. Coronado Villalobos$^{2}$}\email{ccoronado@idu.uff.br}

\affiliation{\\$^1$Universidade Estadual Paulista (Unesp)\\Faculdade de Engenharia, Guaratinguet\'a, Departamento de F\'isica e Qu\'imica\\
12516-410, Guaratinguet\'a, SP, Brazil. \\ $^2$Instituto de Física,
Universidade Federal Fluminense (UFF),\\
24210-346,  Niterói, RJ, Brazil. }


\begin{abstract}
In this present communication we provide a new derivation of the Dirac dual structure by employing a different approach from the originally proposed. Following a general and rigorous mathematical process to compute the dual structure, we investigate if is possible to break the existing ``rigidity'' in its primordial formulation. For this task, firstly, we look towards to understand the core of the Dirac spinors construction \cite{pdirac} and then, we suggest to built an alternative dual structure for the Dirac spinor, which preserve an invariant norm under any $SL(2,\mathcal{C})$ transformation. Finally, we verify if the prominent physical contents are maintained or if it is affected by such construction.
\end{abstract}

\maketitle

\section{Introduction}

Commonly in the textbooks are presented the two problems that the Klein-Gordon equation suffer: the probability density not positive definite, and states of negative energy that occur. These problems arise due to the interpretation of Klein-Gordon equation as a single particle equation of a given wave-function. Dirac's insight was to look towards a first-order equation (field equation), aiming to circumvent the mentioned problems. The solutions of Dirac equation, namely Dirac spinors, cannot be interpreted as wave functions due to the fact that they do not transform from one frame to another unitarily, for this reason we are led to the second quantization (which has an interpretation as a many particle theory), and then we arrive at a successful particle interpretation and the problems associated with the Klein-Gordon equation are bypassed.

As it is well known, much of the physics associated to spinor fields is unveiled from its bilinear covariants for the simple reason that single fermions are not directly experienced \cite{bilineares}. As an useful exercise to the reader, we must remember that a single spinor does not have an usual transformation under a complete rotation, \emph{i.e.}, $\psi\stackrel{2\pi}{\rightarrow} -\psi$, thus, to be physically and mathematically coherent, we need to define a structure that, when combined with the spinor $\psi$, provides $\bar{\psi}\psi\stackrel{2\pi}{\rightarrow}\bar{\psi}\psi$, where $\bar{\psi}$ is called by adjoint, note that the dual along with the usual spinor structure bring us relevant physical information.
So, in view of the foregoing, Dirac solved the problem of the probability density not positive definite, by using the solutions of his first order differential equation, thus, the current density  defined as $J^{\mu}=\bar{\psi}\gamma^{\mu}\psi$ where $\bar{\psi}=\psi^{\dag}\gamma_0$ is the adjoint, provide a probability density positive definite. It should be emphasized that $J^{\mu}$ is not the only physical observable (bilinear covariant), besides the current density we also have, for instance, in the usual case, bearing in mind the relativistic description of the electron: the invariant length $\sigma=\bar{\psi}\psi$, the pseudo scalar amount $\omega=\bar{\psi}\gamma^{5}\psi$, the spin projection in the momentum direction $K^{\mu}=\bar{\psi}\gamma^{\mu}\gamma^{5}\psi$ and the momentum electromagnetic density $S^{\mu\nu}=\bar{\psi}i\gamma^{\mu\nu}\psi$ \cite{crawford1}. In such a way, we clearly see one of the needs of defining the adjoint structure. 

Motivated by the impact of Dirac formulation and the recently new theoretical spin $1/2$ particle endowed with mass-dimension-one, which hold a new adjoint structure contrasting with the aforementioned case, we decided to verify the fundamental requirements for the adjoint construction and also we intend to show if there exists an inherent rigidity or whether it is possible to obtain new structure for the Dirac adjoint. Following a mathematical protocol, thus, we reach to a new dual structure which hold the same physical content as the previous adjoint definition, however, as a price to be paid, such adjoint structure is ``helicity dependent'', as it can be seen in the scope of this work. This paper is organized as is follows: Section \ref{genesis} a brief overview about the theory of Dirac spinors is introduced. Then, in the Section \ref{capdual}, we introduce a general prescription to built the dual structure for any spinor, thus, following this recipe we define the new possible dual structure, elucidating the contrasting subtleties. Section \ref{capfisica} is reserved to explore the physical content of the new structure. Lastly, in Section \ref{conclusion} we conclude.

\section{Genesis: Basic definitions and a brief overview}\label{genesis}
We start our exposition by defining the Dirac spinor as it is commonly presented in the current literature \cite{ryder}
\begin{eqnarray}\label{eq1}
\psi(\boldsymbol{p})= \left(\begin{array}{c}
\phi_{R}(\boldsymbol{p}) \\ 
\phi_{L}(\boldsymbol{p})
\end{array} \right),
\end{eqnarray}
the right and left-handed Weyl spinors transforms under a Lorentz boost like
\begin{eqnarray}
\label{maodir}\phi_{R}(\boldsymbol{p})=e^{+\vec{\sigma}\cdot\hat{\varphi}/2}\phi_{R}(\boldsymbol{0}), \\
\label{maoesq}\phi_{L}(\boldsymbol{p})=e^{-\vec{\sigma}\cdot\hat{\varphi}/2}\phi_{L}(\boldsymbol{0}),
\end{eqnarray}
yielding a boost parameter given by $\cosh(\varphi)=E/m$, $\sinh(\varphi)=p/m$ and $\hat{\varphi}=\hat{p}$.

The spinor presented in \eqref{eq1} is an irreducible representation of the Lorentz group with parity symmetry playing a crucial role of linking the $(1/2,0)\oplus(0,1/2)$ representation spaces, as it can be seen in details in References \cite{ryder,speranca}. For a direct sum of the representation spaces, the boost generators reads
\begin{eqnarray}\label{boostkappa}
\kappa = \left(\begin{array}{cc}
-i\boldsymbol{\sigma}/2 & 0 \\ 
0 & +i\boldsymbol{\sigma}/2
\end{array} \right),
\end{eqnarray}
such boost generator allow us to connect the rest spinor $\psi(\boldsymbol{0})$ with one described by an arbitrary momentum $\psi(\boldsymbol{p})$, as it can be seen
\begin{eqnarray}\label{spinorboostado}
\psi(\boldsymbol{p}) = \exp(i\kappa\cdot\hat{\varphi})\psi(\boldsymbol{0}),
\end{eqnarray}
with the generators of boost at hands, we can readily evaluate the boost operator
\begin{eqnarray}
\exp(i\kappa\cdot\hat{\varphi}) = \left(\begin{array}{cc}
e^{\boldsymbol{\sigma}/2\cdot\hat{\varphi}} & 0 \\ 
0 & e^{-\boldsymbol{\sigma}/2\cdot\hat{\varphi}}
\end{array} \right)\\
=\sqrt{\frac{E+m}{2m}}\left(\begin{array}{cc}\label{lt7}
\mathbbm{1}+\frac{\vec{\sigma}\cdot\vec{p}}{E+m} & 0 \\ 
0 & \mathbbm{1}-\frac{\vec{\sigma}\cdot\vec{p}}{E+m}
\end{array} \right),
\end{eqnarray}
thus, the Lorentz transformed single helicity spinor reads
\begin{eqnarray}
\psi(\boldsymbol{p})=\sqrt{\frac{E+m}{2m}}\left(\begin{array}{c}
\big(\mathbbm{1}+\frac{p\vec{\sigma}\cdot\hat{p}}{E+m}\big)\phi_{R}(\boldsymbol{0}) \\ 
\big(\mathbbm{1}-\frac{p\vec{\sigma}\cdot\hat{p}}{E+m}\big)\phi_{L}(\boldsymbol{0})
\end{array} \right).
\end{eqnarray}
Suppose, the components in equation \eqref{maodir} and \eqref{maoesq}, under action of the parity symmetry, as a consequence we experience the following relation: $\phi_{R}(\boldsymbol{p})\leftrightarrow\phi_{L}(\boldsymbol{p})$, due to the fact that the boost generator change sign, 
so, the previous statement turns possible to write 
\begin{eqnarray}
&&e^{+\vec{\sigma}\cdot\hat{\varphi}/2}\phi_{R}(\boldsymbol{0}) \rightarrow e^{-\vec{\sigma}\cdot\hat{\varphi}/2}\phi_{R}(\boldsymbol{0})=\phi_{L}(\boldsymbol{p}), \\
&&e^{-\vec{\sigma}\cdot\hat{\varphi}/2}\phi_{L}(\boldsymbol{0}) \rightarrow e^{+\vec{\sigma}\cdot\hat{\varphi}/2}\phi_{L}(\boldsymbol{0})=\phi_{R}(\boldsymbol{p}).
\end{eqnarray}
Looking towards a spinor which satisfy the Dirac equation, we impose to the right and left hand components to be eigenstates of the helicity operator, $\vec{\sigma}\cdot\hat{p}$, satisfying
\begin{eqnarray}\label{helicidader}
\vec{\sigma}\cdot\hat{p}\phi^{\pm}_{R}(\boldsymbol{0})=\pm\phi^{\pm}_{R}(\boldsymbol{0}),
\end{eqnarray}
and
\begin{eqnarray}\label{helicidadel}
\vec{\sigma}\cdot\hat{p}\phi^{\pm}_{L}(\boldsymbol{0})=\pm\phi^{\pm}_{L}(\boldsymbol{0}),
\end{eqnarray}
where the four-momentum vector is given by $p^{\mu}=(E, p\sin(\theta)\cos(\phi), p\sin(\theta)\sin(\phi), p\cos(\theta))$. Through the relations above, it becomes easy to define the rest frame spinors up to a phase
\begin{eqnarray}\label{compmais}
\phi^{+}_{R}(\boldsymbol{0})= \phi^{+}_{L}(\boldsymbol{0}) = \sqrt{m}e^{i\varpi_1}\left(\begin{array}{c}
\cos(\theta/2) e^{-i\phi/2} \\ 
\sin(\theta/2) e^{i\phi/2}
\end{array} \right), 
\end{eqnarray}
and
\begin{eqnarray}\label{compmenos}
 \phi^{-}_{R}(\boldsymbol{0})=\phi^{-}_{L}(\boldsymbol{0}) = \sqrt{m}e^{i\varpi_2}\left(\begin{array}{c}
-\sin(\theta/2) e^{-i\phi/2} \\ 
\cos(\theta/2) e^{i\phi/2}
\end{array} \right), 
\end{eqnarray}
in this work we take $\varpi_1=\varpi_2=0$.
Then, applying the Lorentz transformation \eqref{lt7} in the components given in \eqref{compmais} and \eqref{compmenos}, immediately follows that the transformed one can be written as
\begin{eqnarray}
\phi_R^{\pm}(\boldsymbol{p}) = \frac{E+m+\vec{\sigma}\cdot\vec{p}}{[2m(E+m)]^{1/2}}\phi_R^{\pm}(\boldsymbol{0}),
\end{eqnarray}
analogously for the left handed component
\begin{eqnarray}
\phi_L^{\pm}(\boldsymbol{p}) = \frac{E+m-\vec{\sigma}\cdot\vec{p}}{[2m(E+m)]^{1/2}}\phi_L^{\pm}(\boldsymbol{0}).
\end{eqnarray}
As remarked in \cite{nogo}, an important fact concerning to the relations in the rest-frame referential must be highlighted, the derivation of the Dirac equation is only completed correct accomplished only if one assume $\phi^{\pm}_{R}(\boldsymbol{0}) = \phi^{\pm}_{L}(\boldsymbol{0})$ for particles and do not neglect $\phi^{\pm}_{R}(\boldsymbol{0}) = -\phi^{\pm}_{L}(\boldsymbol{0})$ for antiparticles, as commonly neglected in the most of the textbooks. 

Taking advantage of equations \eqref{helicidader} and \eqref{helicidadel}, we find the following set of single helicity spinors 
\begin{eqnarray}\label{espinoresdirac}
\psi_{1}(\boldsymbol{p})=\sqrt{\frac{E+m}{2m}}\left(\begin{array}{c}
\big(\mathbbm{1}+\frac{p\vec{\sigma}\cdot\hat{p}}{E+m}\big)\phi^{+}_{R}(\boldsymbol{0}) \\ 
\big(\mathbbm{1}-\frac{p\vec{\sigma}\cdot\hat{p}}{E+m}\big)\phi^{+}_{L}(\boldsymbol{0})
\end{array} \right),
\\ 
 \psi_{2}(\boldsymbol{p})=\sqrt{\frac{E+m}{2m}}\left(\begin{array}{c}
\big(\mathbbm{1}+\frac{p\vec{\sigma}\cdot\hat{p}}{E+m}\big)\phi^{-}_{R}(\boldsymbol{0}) \\ 
\big(\mathbbm{1}-\frac{p\vec{\sigma}\cdot\hat{p}}{E+m}\big)\phi^{-}_{L}(\boldsymbol{0})
\end{array} \right).
\end{eqnarray}

Looking towards a complete set of eigenspinors of parity operator, we impose to $\psi_{1}(\boldsymbol{p})$ and $\psi_{2}(\boldsymbol{p})$ to satisfy the parity relation $\mathcal{P}\psi(\boldsymbol{p})=\pm\psi(\boldsymbol{p})$, as firstly observed and developed in \cite{speranca}, the operator $\mathcal{P}$ reads  
\begin{eqnarray}
\mathcal{P} = m^{-1}\gamma_{\mu}p^{\mu}.
\end{eqnarray}
Regardless of, we stablish a complete set of eigenspinors of parity operator  
\begin{eqnarray}
&&\label{dirac1}\psi_{1}(\boldsymbol{p})=\left(\begin{array}{c}
\Upsilon_{+}(p^{\mu})\phi_{R}^{+}(\boldsymbol{0}) \\ 
\Upsilon_{-}(p^{\mu})\phi_{L}^{+}(\boldsymbol{0})
\end{array} \right), \\ 
&&\label{dirac2}\psi_{2}(\boldsymbol{p})=-\left(\begin{array}{c}
\Upsilon_{-}(p^{\mu})\phi_{R}^{-}(\boldsymbol{0}) \\ 
\Upsilon_{+}(p^{\mu})\phi_{L}^{-}(\boldsymbol{0}) 
\end{array} \right), \\ 
&&\label{dirac3}\psi_{3}(\boldsymbol{p})=-\left(\begin{array}{c}
\Upsilon_{+}(p^{\mu})\phi_{R}^{+}(\boldsymbol{0}) \\ 
\Upsilon_{-}(p^{\mu})\phi_{L}^{+}(\boldsymbol{0}) 
\end{array} \right), \\ 
&&\label{dirac4}\psi_{4}(\boldsymbol{p})=\left(\begin{array}{c}
\Upsilon_{-}(p^{\mu})\phi_{R}^{-}(\boldsymbol{0}) \\ 
\Upsilon_{+}(p^{\mu})\phi_{L}^{-}(\boldsymbol{0}) 
\end{array} \right),
\end{eqnarray}
in order to summarize the notation, we chosen to express the boost factor as
\begin{eqnarray}
\Upsilon_{\pm}(p^{\mu}) = \sqrt{\frac{E+m}{2m}}\bigg(1\pm\frac{p}{E+m}\bigg),
\end{eqnarray}
such spinors are restricted to the following condition 
\begin{eqnarray}
\mathcal{P}\psi_{i}(\boldsymbol{p}) = +\psi_{i}(\boldsymbol{p}), \quad\mbox{for $i=1$ and $i=2$}, \\
\mathcal{P}\psi_{i}(\boldsymbol{p}) = -\psi_{i}(\boldsymbol{p}), \quad\mbox{for $i=3$ and $i=4$},
\end{eqnarray}
in view of this observation, we state the following
\begin{eqnarray}
(\gamma_{\mu}p^{\mu}-m\mathbbm{1})\psi_{i}(\boldsymbol{p})=0, \quad\mbox{for $i=1$ and $i=2$}, 
\\
(\gamma_{\mu}p^{\mu}+m\mathbbm{1})\psi_{i}(\boldsymbol{p})=0, \quad\mbox{for $i=3$ and $i=4$}, 
\end{eqnarray}
leading us to conclude that $\psi_{1}(\boldsymbol{p})$ and $\psi_{3}(\boldsymbol{p})$ are related with particles, and the remaining two stands for antiparticle. Thus, we have defined a complete set of eigenspinors of parity operator governed by the Dirac dynamic.

\section{The emergence of a new dual structure}\label{capdual}
As previously mentioned, the program to be implemented and accomplished in the scope of the present work is to analyse the possibility to find a new approach to the dual formulation of the Dirac spinor. Such an approach is taken by the authors as important because recently a new class of spin $1/2$ mass-dimension-one dual helicity fermionic field was theoretically discovered and was introduced to the literature, see Ref\cite{1305}, and it shows a new dual structure, contrasting with the previous existing cases, e.g., Majorana and Dirac. The mass-dimension-one fields has a dual structure slightly different, it reads 
\begin{eqnarray}\label{dualelko}
\stackrel{\neg}{\lambda}_{h}(\boldsymbol{p}) = [\Xi(\boldsymbol{p})\lambda_{h}(\boldsymbol{p})]^{\dag}\gamma_0,
\end{eqnarray}
where $h$ stands for the dual helicity, the operator $\Xi(\boldsymbol{p})$ is indeed to exist, $\Xi^2(\boldsymbol{p})=\mathbbm{1}$ ensuring a invertible map, and last but not least, it is required its action on any $\lambda_{h}(\boldsymbol{p})$ yields one of the spinors $\lambda_{h^{\prime}}(\boldsymbol{p})$ from the existing set. Once such requirements are met, this new structure yields a real, non-null and invariant norm for the mass-dimension-one fermionic field.
Faced with this, it leads us to search for a possible new structure for Dirac's spinor. The usual formulation of Dirac spinors is well known to lead to a dual structure given by
\begin{eqnarray}\label{diracdual}
\bar{\psi}(\boldsymbol{p})  = [\mathbbm{1}\psi(\boldsymbol{p})]^{\dag}\gamma_0.
\end{eqnarray}
The questions that arise and remains open are the following: Does the structure presented in \eqref{dualelko} fit into Dirac's case? Is the Dirac structure unique? In view of this facts, we are looking for a new Dirac dual structure based on something similarly to \eqref{dualelko}. Henceforth, we will define it as it reads\footnote{We choose to define the operator $M(\boldsymbol{p})$ rather then $\Xi(\boldsymbol{p})$ given to the fact of avoiding confusion and since it does not necessarily holds the same structure as $\Xi(\boldsymbol{p})$.}
\begin{eqnarray}\label{newdiracdual}
\stackrel{\neg}{\psi}_{i}(\boldsymbol{p}) \stackrel{def}{=} [M(\boldsymbol{p})\psi_{i}(\boldsymbol{p})]^{\dag}\eta,
\end{eqnarray}
where $M(\boldsymbol{p})$ stands for an arbitrary $4\times 4$ matrix and $\eta$ will be posteriorly fixed. Note that the main difference between the dual structure given in \eqref{diracdual} and \eqref{newdiracdual} is the replacement of the identity matrix by $M(\boldsymbol{p})$. Using this prescription, we now evaluate the dual structure for the Dirac spinors given in Equations \eqref{dirac1}-\eqref{dirac4}. The requirement of a Lorentz invariant norm  translates to the assertion that $\eta$ in \eqref{newdiracdual} must to anti-commute with the generators
of boosts ($\kappa$) given in \eqref{boostkappa}, and commute with the generators of the rotations ($\zeta$), given by 
\begin{eqnarray}
\zeta= \left(\begin{array}{cc}
\boldsymbol{\sigma}/2 & 0 \\ 
0 & \boldsymbol{\sigma}/2
\end{array} \right),
\end{eqnarray}
thus,
\begin{eqnarray}\label{releta}
\{\kappa_{i},\eta \}= 0, \quad [\zeta_{i}, \eta]=0,
\end{eqnarray}
where $i=x, y, z$, a straight forward calculation obeying the constraints in \eqref{releta} result in $\eta=\gamma_0$, in agreement with previously stated in a general context of spinor in \cite{plb687, prd83}. 

To illustrate the mathematical procedure to evaluate $\stackrel{\neg}{\psi}(\boldsymbol{p})$, suppose a spinor corresponding to a particle, then it follows\footnote{Authors remark that the same mathematical procedure can be employed to compute the dual structure for antiparticle spinors, however, one can not neglect the relation  $\phi_{R}(\boldsymbol{0})= -\phi_{L}(\boldsymbol{0})$, for this specific case.}
\begin{eqnarray}
\phi_{R}(\boldsymbol{0}) = \phi_{L}(\boldsymbol{0})=  \left(\begin{array}{c}
a \\ 
b
\end{array} \right),
\end{eqnarray}
and the operator $M(\boldsymbol{p})$ reads
\begin{eqnarray}\label{mm}
M(\boldsymbol{p}) = \left(\begin{array}{cc}
M_{11} & M_{12} \\ 
M_{21} & M_{22}
\end{array} \right),
\end{eqnarray}
where $M_{ij}$ stands for a $2\times 2$ matrix composed by $w_{ij}$ elements. Thus, the norm is, in abstract, given by
\begin{eqnarray*}
\stackrel{\neg}{\psi}(\boldsymbol{p})\psi(\boldsymbol{p})= \bigg\lbrace&\frac{(E+p)}{m}&(|a|^2w^*_{31}+ab^*w^*_{32}+a^*bw^*_{41}+|b|^2w^*_{42}) 
\\\nonumber
+&\frac{(E-p)}{m}&(|a|^2w^*_{13}+ab^*w^*_{14}+a^*bw^*_{23}+|b|^2w^*_{24})
\\\nonumber
+&&(|a|^2w^*_{33}+ab^*w^*_{34}+a^*bw^*_{43}+|b|^2w^*_{44})
\\\nonumber
+&&(|a|^2w^*_{11}+ab^*w^*_{12}+a^*bw^*_{21}+|b|^2w^*_{22})
\bigg\rbrace,
\end{eqnarray*}
imposing $\stackrel{\neg}{\psi}(\boldsymbol{p})\psi(\boldsymbol{p})$ to be Lorentz invariant, if we choose to recover the original prescription, the only non-null terms must be $w_{ij}$ for $i=j$, and then we have $M(\boldsymbol{p})= \mathbbm{1}$, note that in this case the only remaining components are square-like functions. Nevertheless, our intention is to look towards an alternative method, in such a way, we take advantage of the square-like components that is not on its main diagonal. Regardless of, the dual by construction allows two possibilities, it thus reads
\begin{eqnarray}\label{espinormais}
\stackrel{\neg}{\psi}_{i}(\boldsymbol{p}) = [M_{+}(\boldsymbol{p})\psi_{i}(\boldsymbol{p})]^{\dag}\gamma_0, \quad\mbox{for $i=1$ and $i=3$},
\end{eqnarray}
and
\begin{eqnarray}\label{espinormenos}
\stackrel{\neg}{\psi}_{i}(\boldsymbol{p}) = [M_{-}(\boldsymbol{p})\psi_{i}(\boldsymbol{p})]^{\dag}\gamma_0, \quad\mbox{for $i=2$ and $i=4$},
\end{eqnarray}
the presence of the lower indexes $+$ and $-$ are related with the helicity of the components $\phi_{R/L}$, and it follows that $M(\boldsymbol{p})$ operator in the matricial form reads
\begin{eqnarray}\label{mmais}
M_{+}(\boldsymbol{p}) = \left(\begin{array}{cccc}
0 & 0 & \frac{m}{E-p} & 0 \\ 
0 & 0 & 0 & \frac{m}{E-p} \\ 
\frac{m}{E+p} & 0 & 0 & 0 \\ 
0 & \frac{m}{E+p} & 0 & 0
\end{array} \right), 
\end{eqnarray}
and
\begin{eqnarray}\label{mmenos}
 M_{-}(\boldsymbol{p}) = \left(\begin{array}{cccc}
0 & 0 & \frac{m}{E+p} & 0 \\ 
0 & 0 & 0 & \frac{m}{E+p} \\ 
\frac{m}{E-p} & 0 & 0 & 0 \\ 
0 & \frac{m}{E-p} & 0 & 0
\end{array} \right), 
\end{eqnarray}
both matrices obey the following properties $M_{\pm}^{2}(\boldsymbol{p}) = \mathbbm{1}$ and $M^{-1}_{\pm}(\boldsymbol{p})$ exist. As main result, note that for this new construction the dual structure is not unique, which strongly contrast with the previous case. It should be mentioned that the dual structure depends on the helicity of the components, as it can be seen in equations \eqref{espinormais} and \eqref{espinormenos}; spinors endowed with positive helicity components  have an operator $M(\boldsymbol{p})$ different from the one endowed with negative helicity components, this is the price to be paid.

Such new structure provide us a real and non-null Lorentz invariant norm. Nevertheless, we must point out a fact that has already been discussed, the action of \eqref{mmais} and \eqref{mmenos} on the $\psi(\boldsymbol{p})$ yields a spinor from the established set, in other words,  
\begin{eqnarray}
&&M_{+}(\boldsymbol{p})\psi_{i}(\boldsymbol{p})=\psi_{i}(\boldsymbol{p}), \quad \mbox{for $i=1$ and $i=3$},\\
&&M_{-}(\boldsymbol{p})\psi_{i}(\boldsymbol{p})=\psi_{i}(\boldsymbol{p}), \quad \mbox{for $i=2$ and $i=4$},
\end{eqnarray}
in this case the matrix $M_{+}(\boldsymbol{p})$ and $M_{-}(\boldsymbol{p})$ acting on the spinors behaves like identity matrix. 

So, now we are able to write the dual structures in terms of $M_{\pm}(\boldsymbol{p})$, so it reads
\begin{eqnarray}
\stackrel{\neg}{\psi}_{1}(\boldsymbol{p})&=&\left(\begin{array}{cc}
\Upsilon_{-}(p^{\mu})\phi^{+\dag}_{R}(\boldsymbol{0}) & \Upsilon_{+}(p^{\mu})\phi^{+\dag}_{L}(\boldsymbol{0})
\end{array} \right),\\
\stackrel{\neg}{\psi}_{2}(\boldsymbol{p})&=& -\left(\begin{array}{cc}
\Upsilon_{+}(p^{\mu})\phi^{-\dag}_{R}(\boldsymbol{0}) & \Upsilon_{-}(p^{\mu})\phi^{-\dag}_{L}(\boldsymbol{0})
\end{array} \right),  \\
\stackrel{\neg}{\psi}_{3}(\boldsymbol{p})&=&\left(\begin{array}{cc}
\Upsilon_{-}(p^{\mu})\phi^{+\dag}_{R}(\boldsymbol{0}) & \Upsilon_{+}(p^{\mu})\phi^{+\dag}_{L}(\boldsymbol{0})
\end{array} \right), \\
\stackrel{\neg}{\psi}_{4}(\boldsymbol{p})&=&-\left(\begin{array}{cc}
\Upsilon_{+}(p^{\mu})\phi^{-\dag}_{R}(\boldsymbol{0}) & \Upsilon_{-}(p^{\mu})\phi^{-\dag}_{L}(\boldsymbol{0})
\end{array} \right),
\end{eqnarray}
thus, we accomplished the protocol of verifying the possibility of obtaining new dual structures, in addition, we have seen that it is possible and without loss of generality we have obtained analogous structures as previously defined. Next step is to analyse its physical contents.

\section{The physics behind the new dual structure}\label{capfisica}

This section is reserved to analyse the physical contents of the new dual structure. The above structures satisfies the orthonormal relations (by construction), so, we have
\begin{eqnarray}
\stackrel{\neg}{\psi}_{i}(\boldsymbol{p})\psi_{i}(\boldsymbol{p})=+2m, \quad\mbox{for $i=1$ and $i=2$}, \\
\stackrel{\neg}{\psi}_{i}(\boldsymbol{p})\psi_{i}(\boldsymbol{p})=-2m, \quad \mbox{for $i=3$ and $i=4$}.
\end{eqnarray}
Insomuch, it is important to evaluate the spin sums once we found a new dual structure, and then verify if such physical amount    
is proportional or unitarily connected to the Dirac momentum–space wave operator, a slightly lengthy calculation reveal us the following relations  
\begin{eqnarray}
&&\sum_{i=1,2}\psi_{i}(\boldsymbol{p})\stackrel{\neg}{\psi}_{i}(\boldsymbol{p})=(\gamma_{\mu}p^{\mu}+m\mathbbm{1}),\\
&&\sum_{i=3,4}\psi_{i}(\boldsymbol{p})\stackrel{\neg}{\psi}_{i}(\boldsymbol{p})=(\gamma_{\mu}p^{\mu}-m\mathbbm{1}).
\end{eqnarray}
As a consequence the main result for the new Dirac formulation shows that the spin sums found determines not only the wave operator but also determines the structure of the Feynman–Dyson propagator, leading us to conclude that it matches with the  previous established amounts, both formulations are indistinguishable, in other words, the new dual does not affect in any level the fermionic statistics neither the propagator structure. Going further, if one evaluate the physical observables, encoded by the 16 bilinear covariants, given by
\begin{eqnarray}
\label{covariantes}
&&\sigma=\psi^{\dag}\gamma_{0}\psi, \;\;  \omega=-\psi^{\dag}\gamma_{0}\gamma_{0123}\psi, \;\; J_\mu=\psi^{\dag}\mathrm{\gamma_{0}}\gamma_{\mu}\psi, \nonumber
\\
&&K_\mu=\psi^{\dag}\mathrm{\gamma_{0}}\textit{i}\mathrm{\gamma_{0123}}\gamma_{\mu}\psi,\;\; S_{\mu\nu}=\frac{1}{2}\psi^{\dag}\mathrm{\gamma_{0}}i\gamma_{\mu\nu}\psi,
\end{eqnarray}
one find exactly the same results already found in \cite{rcd}, regardless of, even replacing $\psi^{\dag}\gamma_{0}$ by $[M(\boldsymbol{p})\psi]^{\dag}\gamma_{0}$ in the bilinear forms given in \eqref{covariantes}, the result remains unchanged, such spinor belongs to type-2 within Lounesto classification. This leads us to think that regardless of the dual structure chosen, the physical information contained in the spinors is unchanged. 

\section{Final Remarks}\label{conclusion}
The purpose of this work, was to show an alternative approach to construct the Dirac dual structure, following the same program as it was recently made for the mass-dimension-one fields proposed in the literature.
So, we have shown the complete mathematical protocol to compute the dual structure for any spinor, such program is given by a simple recipe as it was shown in details in Section \ref{capdual}. 
Although the way it is commonly presented, the Dirac dual structure is not unique. We have seen in the scope of this paper, we can move towards for a more general and broad approach and build a different dual structure, which in turn, is ``helicity dependent'', we suppose that is the price to be paid once we are leaving the original formulation. However, physically and mathematically, such new structure encodes all the same physical information as the previous one.

\section{Acknowledgments}
Authors express their gratitude to Professor Dharam Vir Ahluwalia for careful reading the entire first draft of the manuscript and provided many insightful suggestions. We are also grateful to Professor Julio Marny Hoff da Silva for his hopeful questions and the ensuing discussions.
RJBR thanks to CAPES for the financial support and CHCV thanks to PNPD-CAPES for the financial support.

\end{document}